\pdfoutput=1
\documentclass[11pt]{article}

\usepackage[utf8]{inputenc}
\usepackage[T1]{fontenc}
\usepackage{textcomp}
\usepackage{lmodern}
\usepackage{graphicx}
\usepackage{booktabs}
\usepackage{amsmath}
\usepackage{amssymb}
\usepackage[hidelinks]{hyperref}
\usepackage[margin=1in]{geometry}
\usepackage{microtype}
\DeclareUnicodeCharacter{00B0}{\textdegree}
\DeclareUnicodeCharacter{2009}{\,}
\DeclareUnicodeCharacter{2010}{-}
\DeclareUnicodeCharacter{2013}{\textendash}
\DeclareUnicodeCharacter{2212}{\ensuremath{-}}
\DeclareUnicodeCharacter{2082}{\ensuremath{_2}}
\DeclareUnicodeCharacter{2083}{\ensuremath{_3}}
\DeclareUnicodeCharacter{03B1}{\ensuremath{\alpha}}
\DeclareUnicodeCharacter{03B2}{\ensuremath{\beta}}
\DeclareUnicodeCharacter{0307}{}

\usepackage[backend=biber,style=numeric-comp,sorting=none,
            giveninits=true,doi=true,isbn=false,url=false]{biblatex}
\title{\textbf{Engineering strategies for high-performance self-powered
zinc oxide (ZnO) ultraviolet photodetectors: a design-oriented review}}
\author{Eka Nurfani$^{\,a,b,}$\thanks{Corresponding author. \emph{E-mail address:}
eka.nurfani@mt.itera.ac.id}, \ Keysya C. A. Putri$^{\,a}$, \
Tarmizi Taher$^{\,b,c}$, \\[3pt]
Aditya Rianjanu$^{\,a,b}$ \ and \ Muhamad F. Arif$^{\,a,b}$ \\[6pt]
\small $^{a}$ Department of Materials Engineering, Faculty of Industrial
Technology, \\
\small Institut Teknologi Sumatera (ITERA), South Lampung 35365, Indonesia \\[1pt]
\small $^{b}$ Center for Green and Sustainable Materials, \\
\small Institut Teknologi Sumatera (ITERA), South Lampung 35365, Indonesia \\[1pt]
\small $^{c}$ Department of Environmental Engineering, \\
\small Institut Teknologi Sumatera (ITERA), South Lampung 35365, Indonesia}
\date{}

\begin{document}
\maketitle

\begin{abstract}
\noindent Self-powered ultraviolet (UV) photodetectors generate a signal at zero
applied bias, removing standby power and suppressing dark current, suiting
battery-free Internet-of-Things sensing. Zinc oxide (ZnO) is the archetypal
absorber for this task: UV-selective, earth-abundant, non-toxic, and processable
at low temperature from solution. Operating it without bias, however, requires a
built-in asymmetry that pristine ZnO rarely provides. Because the literature is
organized by material or mechanism, the design logic remains obscured. Here we
organize the field by \emph{engineering strategy}, identifying five families of
levers (junction and built-in-field engineering, doping and defect control,
polarization coupling, interface/surface and photoelectrochemical engineering, and
architecture and light management) and mapping each onto the metrics it serves and
the trade-offs it imposes. A unifying lens is the \emph{zero-bias doping paradox}:
raising carrier density narrows the depletion region that zero-bias collection
needs, so the self-powered optimum differs from the biased one. We flag the
detectivity, speed, and active-area pitfalls that make zero-bias performance easy
to overstate, and condense the synthesis into a decision tree linking target
metric to strategy and growth route. The remaining bottlenecks (persistent
photoconductivity, p-type ZnO, and reporting consistency) are problems of defect
control and rigorous measurement, not of new materials.
\end{abstract}


\section{Introduction}\label{sec:intro}

Ultraviolet (UV) photodetection underpins applications from flame sensing and
water sterilization to environmental monitoring, missile-plume warning, and optical
wireless communication~\cite{xiusheng2021semiconductor, jebas2026progress}.
As these functions migrate into
distributed, battery-constrained platforms (Internet-of-Things nodes, wearables,
and off-grid sensors), detectors that operate without an external power supply
have become increasingly attractive, since removing the battery removes
both a maintenance burden and an environmental one. A \emph{self-powered}
photodetector generates a measurable photoresponse at zero applied bias, drawing
on an internal built-in field rather than a battery, and thereby eliminates
standby power, simplifies array wiring, and suppresses the dark current that
limits low-light detection~\cite{tian2017nanoscale, su2017driven, wei2021organic}. Among the
wide-bandgap metal oxides (ZnO, Ga$_2$O$_3$~\cite{liu2023review}, SnO$_2$, TiO$_2$, and their
relatives), zinc oxide (ZnO) is the archetypal and most widely studied
absorber: intrinsically UV-selective, earth-abundant, non-toxic, chemically
robust, and accessible through inexpensive, low-temperature solution and vapour
routes~\cite{boruah2019zinc, xiusheng2021semiconductor}, a
combination as attractive on sustainability grounds as on optical ones; indeed,
fully green, plant-extract-mediated syntheses of ZnO nanostructures are well
established, with the secondary metabolites of the plant extract acting as the
reducing and capping agents~\cite{faisal2021green, saputra2024green}. We take
ZnO as the reference material throughout and draw on wider-gap oxides mainly for
comparison.

The field has expanded rapidly, and existing reviews are organized almost
entirely along two axes: \emph{material-centric}, surveying ZnO, Ga$_2$O$_3$, 2D
materials, or perovskites individually~\cite{boruah2019zinc, ouyang2021nanomaterials, xiusheng2021semiconductor, jebas2026progress, nalwa2020review, tian2017hybrid, veeramalai2021lead, chow2019organic}, or \emph{mechanism-centric},
treating a single effect (the built-in photovoltaic field, the
photoelectrochemical cell, or the pyro-/piezo-phototronic effect) on its
own~\cite{su2017driven, zhou2015overview, kaur2023review, aftab2023emerging, jia2025pyro, fattah2021sensing}. Neither answers the question a practitioner asks:
\emph{given a target (high responsivity, fast response, solar-blind
selectivity, flexibility, or low cost), which strategy should I reach for, and
what will it cost elsewhere?} The strategies that move these metrics cut across
materials and mechanisms, so a catalogue indexed by either leaves the design logic
implicit.

Table~\ref{tab:reviews} makes the gap concrete. Read down the ``organizing
axis'' column, the material-indexed reviews are complete within a material and
silent across materials: a practitioner who has settled on ZnO learns what has
been tried in ZnO~\cite{boruah2019zinc, xiusheng2021semiconductor} but not that
the same interface problem was solved in a Ga$_2$O$_3$ or 2D device, while the
perovskite, MoS$_2$, and organic surveys~\cite{tian2017hybrid, veeramalai2021lead, nalwa2020review, chow2019organic} carry transferable lessons on
polarization coupling, van der Waals interfaces, and flexible form factors that
never reach the oxide literature. The mechanism-indexed reviews have the mirror
limitation: each treats one field source completely~\cite{zhou2015overview, kaur2023review, jia2025pyro, aftab2023emerging} and therefore cannot say when
that source is the right choice, because the comparison that would answer it lies
outside the review's frame. What is missing is not coverage but a common axis on
which the alternatives can be weighed, and the practitioner's question is
inherently comparative.

\begin{table}[htbp]
\centering
\small
\caption{Organizing axis of representative recent reviews of self-powered and UV
photodetection, and the position this review takes. ``Design logic'' asks whether
the review maps a \emph{target metric} onto a recommended strategy, rather than
surveying devices.}
\label{tab:reviews}
\begin{tabular}{@{}p{3.5cm}p{3.0cm}p{4.0cm}p{3.4cm}@{}}
\toprule
Representative reviews & Organizing axis & Scope & Design logic \\
\midrule
ZnO photodetectors~\cite{boruah2019zinc, xiusheng2021semiconductor} &
material & ZnO films and nanowires & implicit \\
Ga$_2$O$_3$ Schottky DUV diodes~\cite{liu2023review} &
material & one wide-gap oxide & implicit \\
MoS$_2$, perovskite, organic detectors~\cite{nalwa2020review, tian2017hybrid, veeramalai2021lead, chow2019organic} &
material & non-oxide absorbers & implicit \\
Nanomaterial surveys~\cite{ouyang2021nanomaterials, jebas2026progress} &
material class & many absorbers & implicit \\
Built-in-field and nanoscale self-powered devices~\cite{su2017driven, tian2017nanoscale, fattah2021sensing} &
mechanism & one field source & partial \\
PEC-type detectors~\cite{zhou2015overview, kaur2023review} &
mechanism & solid/liquid junctions & partial \\
Pyro-/piezo-phototronic devices~\cite{jia2025pyro, aftab2023emerging} &
mechanism & polar absorbers & partial \\
\midrule
\textbf{This review} & \textbf{engineering strategy} &
ZnO as reference platform, wider-gap oxides as benchmarks &
\textbf{explicit: matrix $+$ decision tree} \\
\bottomrule
\end{tabular}
\end{table}

\begin{figure}[htbp]
\centering
\includegraphics[width=\linewidth]{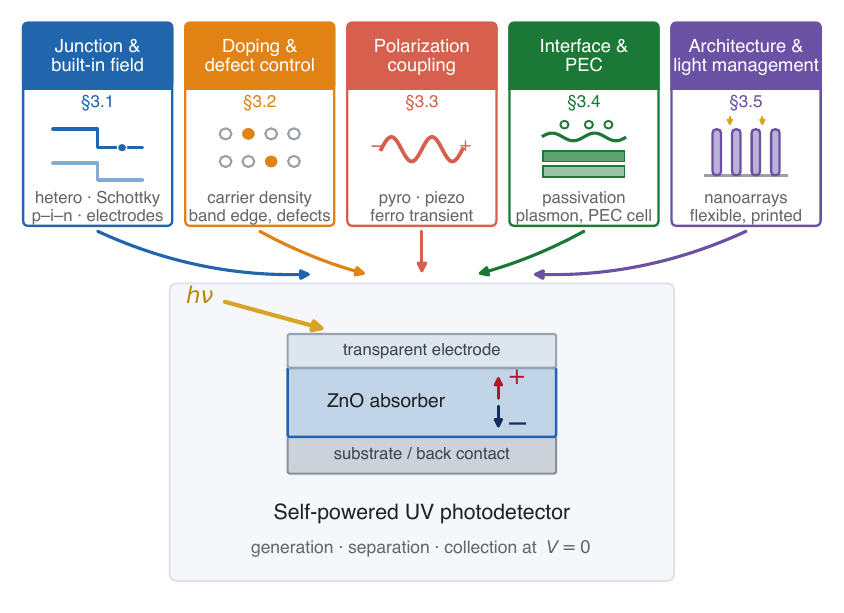}
\caption{The five families of engineering strategy for self-powered UV
photodetectors. Each family is a lever that creates or strengthens the built-in
asymmetry required for charge separation and collection at zero bias; the
strongest devices combine several. Schematic.}
\label{fig:umbrella}
\end{figure}

This review is organized instead by \emph{engineering strategy}, with ZnO as the
reference platform. We identify five families of levers that create or strengthen
the built-in asymmetry a zero-bias device requires: junction and built-in-field
engineering, doping and defect control, polarization coupling, interface/surface
and photoelectrochemical engineering, and architecture and light management
(Fig.~\ref{fig:umbrella}), and we map each onto the target metrics it serves and
the trade-offs it imposes. Three features distinguish the treatment: a single
unifying lens, the \emph{zero-bias doping paradox}, whereby raising carrier
density narrows the depletion region collection depends on ($W\propto N_d^{-1/2}$)
so the self-powered optimum differs fundamentally from the biased one; a
central concern for measurement rigor, flagging the pitfalls that make
zero-bias performance easy to overstate~\cite{wang2023characterize, dearquer2017solution}; and explicit, citable design rules, namely a decision tree
linking target metric to strategy and growth route.

Organizing by strategy has a further payoff: the same physical lever (a clean
van der Waals interface, an intermediate doping optimum, a pyro-phototronic
transient, a solid/liquid potential step) recurs under many material guises, so
a result proven in one system transfers as a \emph{design principle} even when the
materials do not~\cite{su2017driven, boruah2019zinc}. Our taxonomy is therefore
functional, grouping levers by what they do to the built-in asymmetry and cutting
across the material- and mechanism-indexed boundaries of earlier
reviews~\cite{ouyang2021nanomaterials, fattah2021sensing, kaur2023review}.

The scope is deliberately bounded. We focus on ZnO-based UV detection
($\lambda\lesssim420$~nm), where the author's experience lies, and on devices
with an explicit zero-bias (self-powered) claim; wider-gap oxides such as
Ga$_2$O$_3$ (for solar-blind UV-C), together with 2D-material, perovskite, and
photoelectrochemical systems, are drawn in as comparative benchmarks, while biased
photoconductors are cited only for context. We draw on a curated selection of the
recent literature, chosen for mechanistic clarity and completeness of the
reported metrics, rather than an exhaustive census, and all
figures are original, with any replotted data traceable to their cited sources.
Section~\ref{sec:fundamentals} sets out what zero-bias operation demands and how
to measure it rigorously; Section~\ref{sec:strategies} treats the five strategy
families in turn; Section~\ref{sec:compare} compares them through a
strategy--metric matrix and case studies; and Section~\ref{sec:designrules}
presents the resulting design rules and open problems.

\section{What zero-bias operation demands}\label{sec:fundamentals}

Before cataloguing strategies, we isolate the physics that constrains them. Three
ideas recur through every family of Section~\ref{sec:strategies} and are worth
stating first: the built-in asymmetry a zero-bias device cannot do without, the
doping paradox that makes its optimum unlike a biased detector's, and the
measurement pitfalls that make zero-bias performance easy to overstate. These are
less a tutorial than the design constraints every later strategy must satisfy, and
each is stated here only to the depth those strategies later require.

\subsection{Built-in asymmetry as the prerequisite for zero-bias collection}

A photodetector without external power must turn photons into a measurable signal
using only the light's own energy. Under bias, even a symmetric semiconductor
delivers large photocurrent and photoconductive gain, because the applied field
sweeps carriers to the contacts and, in trap-rich oxides, recirculates majority
carriers many times before recombination~\cite{boruah2019zinc, xiusheng2021semiconductor}. At zero bias this driving force vanishes: the only way
to separate an electron--hole pair and deliver it to the circuit is an internal,
\emph{built-in} asymmetry that imposes a preferred direction on carrier
flow~\cite{su2017driven, tian2017nanoscale}. Self-powered UV photodetection is
therefore not one device concept but a family of strategies unified by one
requirement: engineering this asymmetry into an otherwise passive ZnO (or
related wide-gap oxide) film (Fig.~\ref{fig:fund}a).

The asymmetry can be supplied in physically distinct ways: the depletion field
of a p--n/p--i--n junction, the band bending of a Schottky contact, the staggered
alignment of a heterojunction, the work-function difference between two electrodes,
the solid/liquid step of a photoelectrochemical (PEC) cell, or the bound-charge
field of a polar layer~\cite{su2017driven, zhou2015overview, kaur2023review, aftab2023emerging}. However it is created, the physical mechanism at $V=0$
consists of three steps: absorption, separation by the built-in field, and
collection (Fig.~\ref{fig:fund}a). These mechanisms are not interchangeable: a p--n/p--i--n
field is the most reproducible but bounded by the bandgaps of its two constituent
materials; a Schottky barrier is simpler and fast, but limited by a thin depletion layer
and often a low barrier at the metal--semiconductor interface~\cite{liu2023review}; a
semiconductor--semiconductor heterojunction offers a larger offset at the price of
an interface that seeds recombination; and the PEC and polarization routes are
simple and scalable but limited, respectively, by the chemical stability of the
photoelectrode in its operating environment and by the transient nature of the
polarization response (Sections~\ref{sec:interface}, \ref{sec:polarization}).

Whichever mechanism supplies the field, zero-bias collection is governed by where
each photon is absorbed relative to it. Carriers generated inside the depletion (or
bound-charge) region are separated by the built-in field~\cite{su2017driven}; those
in the quasi-neutral bulk must diffuse to the field edge, so collection is
restricted by the minority-carrier diffusion length $L$~\cite{jain2024heterostructured}.
The effective collection depth is thus $\approx W+L$, and with no applied voltage
to widen it, a self-powered detector cannot trade voltage for collection the way a
biased one can. Every
strategy in Section~\ref{sec:strategies} is accordingly an attempt to enlarge
$W+L$, raise $V_\mathrm{bi}$, or place more absorption inside the field.

This constraint has a geometric consequence that decides much of what follows.
Above the band edge, a direct-gap oxide absorbs strongly, so most photons stop
within a shallow surface layer, whereas the depletion region of a moderately
doped film and the diffusion length of a defective, solution-grown oxide are both
short. Whether a device works at zero bias therefore depends on whether these two
short lengths \emph{overlap}, not on either separately: absorption concentrated
outside $W+L$ is converted into recombination rather than photocurrent, and
absorption inside it is collected with near-unity efficiency. Two devices of
identical material and identical absorbance can accordingly differ sharply in
zero-bias response purely through where the field sits relative to the absorption
profile; in a ZnO/Ga$_2$O$_3$ junction, for instance, the depletion region lies
mainly on the Ga$_2$O$_3$ side and the ZnO contributes only a weak
signal~\cite{su2017driven}. This is why the
illumination geometry matters (front- and
back-illumination place the absorption profile on opposite sides of the junction),
why an ultrathin absorber can outperform a thick one, as nanometre-thick amorphous
Ga$_2$O$_3$ on GaN demonstrates~\cite{lin2023nanoscale}, and why nanostructuring,
which decouples the absorption path from the collection
path~\cite{lin2016core, jain2024heterostructured}, recurs in
Section~\ref{sec:architecture} as the one escape from the trade-off available to
a planar film.

The ceiling on $V_\mathrm{bi}$ likewise differs by field source, and this sets a
practical hierarchy. An all-oxide p--n junction between a p-type oxide and a
wide-gap n-type oxide (NiO/Ga$_2$O$_3$~\cite{wang2020oxide} and
NiO/ZnGa$_2$O$_4$~\cite{khan2024unveiling} are the canonical couples) delivers the
largest and most thermodynamically stable step available in an all-oxide stack;
a Schottky contact delivers less, and what it delivers depends on the choice of
metal and on interfacial barrier modulation~\cite{liu2023review}; a
polarization layer adds a bound-charge field that can exceed either, but only
transiently or while the remnant polarization survives; and the solid/liquid step
of a PEC cell is intrinsically large for a wide-gap oxide yet is fixed by the
electrolyte redox potential rather than by the solid. Alloying the host raises the
ceiling for all of them by widening the gap~\cite{zhou2025mgzno}, at the
cost of a phase change at high alloy content discussed in Section~\ref{sec:doping}. A designer thus chooses
not only \emph{how much} built-in potential to build, but how durable and how
tunable the source of that potential should be.

Speed at zero bias is set by a different bottleneck. With no external field,
transit cannot be shortened by raising drift velocity, so the intrinsic limit is
the junction RC time plus transit across $W+L$, both favouring a thin,
strongly-fielded depletion region. In practice an oxide detector is almost always
slower, because trapping at native defects and the persistent photoconductivity
they cause add a long tail that no field shaping
removes~\cite{chen2025spiro, ocaya2023bias}. Zero-bias speed is therefore governed less
by geometry than by the defect landscape, which is why it recurs below as
something doping and interface control, not junction design alone, ultimately
decide.

One consequence frames the rest: the photoconductive gain that dominates biased
oxide detectors is unavailable to a photovoltaic-type detector, whose gain is unity
short of avalanche~\cite{su2017driven}, so the signal is set by how efficiently the
built-in field collects carriers generated within about $W+L$. Any material change reshaping that
field region therefore acts directly on performance, which is precisely where
doping enters, with a counter-intuitive twist.

\subsection{The doping paradox at zero bias}\label{sec:paradox}

\begin{figure}[htbp]
\centering
\includegraphics[width=\linewidth]{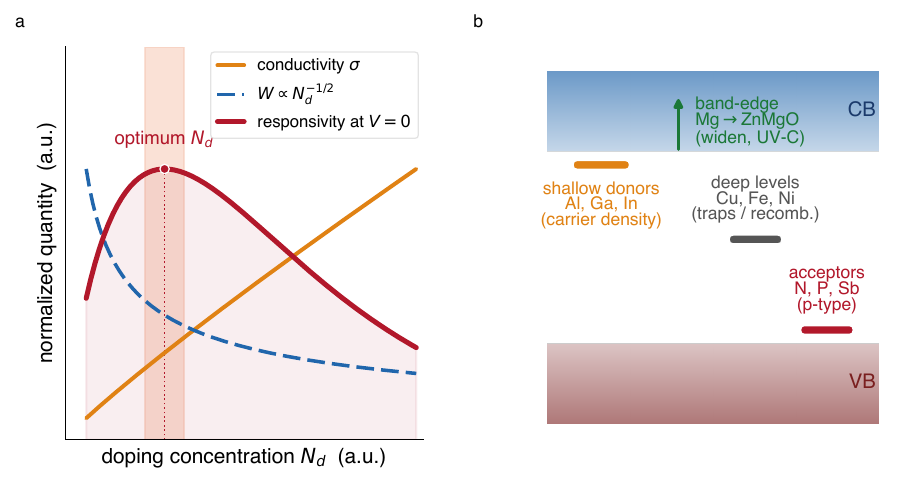}
\caption{The zero-bias doping paradox. (a) As doping $N_d$
rises, conductivity increases while $W\propto N_d^{-1/2}$ shrinks, so the
zero-bias responsivity peaks at an intermediate, optimum $N_d$ (illustrative,
arbitrary units). (b) Dopant function classes in the host-oxide gap: shallow
donors (carrier density), band-edge engineering, deep levels
(traps/recombination), and acceptors (p-type). Both panels schematic; energy
levels not to scale.}
\label{fig:doping}
\end{figure}

Doping is the most accessible lever for tuning an oxide, especially in the
solution-processing laboratories where much self-powered-device work is
done. For a one-sided junction the depletion width obeys
\begin{equation}\label{eq:depletion}
W=\sqrt{\frac{2\varepsilon V_\mathrm{bi}}{qN_d}}\;\propto\;N_d^{-1/2},
\end{equation}
while the peak field at the metallurgical junction grows with $N_d$.
Increasing the donor (or acceptor) density thus raises both conductivity and
the built-in field, yet simultaneously \emph{narrows} the very depletion
region within which zero-bias separation takes place. Under bias this
trade-off is mild, because the external field extends collection well beyond
$W$; at zero bias the collection volume is essentially $W$ plus a diffusion
length, so a film that is doped too heavily becomes highly conductive while
harvesting light from a vanishingly thin layer. The optimum doping level for
a self-powered device is consequently lower, and qualitatively different,
from the optimum for a biased photoconductor (Fig.~\ref{fig:doping}a).

This \emph{doping paradox} predicts a maximum in zero-bias responsivity at an
intermediate concentration (a volcano-shaped dependence on $N_d$), and
reports of doped self-powered oxide detectors repeatedly show such an optimum,
beyond which performance falls as the junction narrows and dopant-related
recombination grows~\cite{wan2021doping, zhang2016development, huang2025boosting}. The
paradox thus establishes doping as a zero-bias-specific design variable, not a
monotonic ``more-is-better'' knob; we examine the element-resolved evidence in
Section~\ref{sec:strategies}.

A maximum exists because three effects compete as $N_d$ rises and only their
product is observable: the built-in field strengthens (aiding separation), the
depletion width narrows as $N_d^{-1/2}$ (shrinking the drift-collection volume),
and ionized-impurity scattering and dopant-related recombination grow (shortening
the lifetime and the diffusion length $L$). The first saturates once the field can
sweep $W$, whereas the second and third worsen without bound, so the responsivity
rises, rolls over, and falls. Two corollaries follow. First, the optimum is not
universal: a strongly absorbed UV-C photon stops within nanometres of the surface
and ignores $W$, whereas a weakly absorbed near-edge photon needs the widest
possible $W+L$, so one film can sit on opposite sides of its optimum across the
spectrum. Second, the optimum can be moved by decoupling doping's two roles, confining heavy doping to a thin contact layer that supplies the field while a
lightly doped absorber keeps $W$ wide, the logic behind the p--i--n and
graded-doping stacks of Section~\ref{sec:junction}.

Native point defects modulate the same balance: in ZnO, oxygen vacancies and hydrogen interstitials act as deep
and shallow donors~\cite{boruah2016energy}, oxygen adsorbed at the surface bends the bands
upward until photogenerated holes desorb it~\cite{boruah2019zinc}, and vacancy
populations seed the persistent photoconductivity that slows the
response~\cite{chen2025spiro}; yet, deliberately engineered, oxygen vacancies can
extend the absorption of a wide-gap ferroelectric crystal into the mid-infrared for
broadband self-powered detection~\cite{wang2025defect}. The practical lever is thus a
dopant--defect ensemble, and the functional taxonomy of
Section~\ref{sec:strategies}, namely carrier-density tuning, band-edge engineering,
deep-level modulation, polarization enhancement, and p-type doping
(Fig.~\ref{fig:doping}b), is organized around these roles, not the periodic
table.

\subsection{Reading zero-bias performance rigorously}\label{sec:metrics}

\begin{figure}[htbp]
\centering
\includegraphics[width=\linewidth]{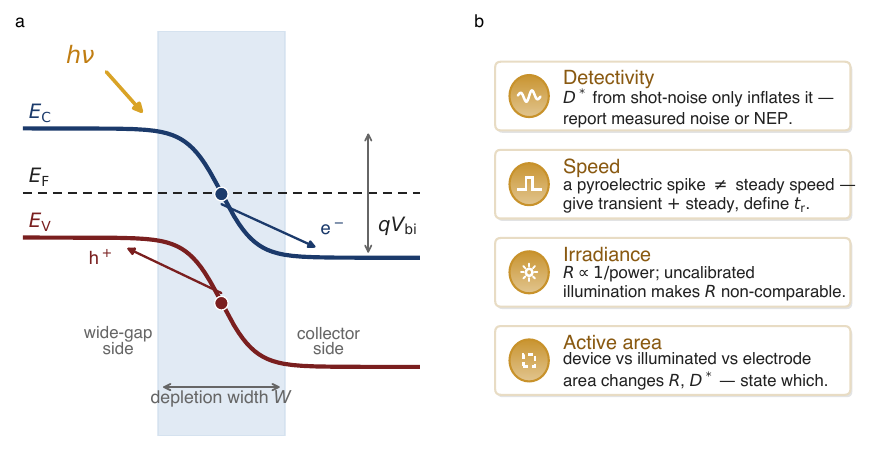}
\caption{What zero-bias operation demands. (a) Band diagram of a
representative built-in field at $V=0$: photogenerated carriers are separated
across the depletion width $W$ by the built-in potential $qV_\mathrm{bi}$ and
collected without external bias. (b) Reporting metrics rigorously at $0$~V:
detectivity, speed, irradiance, and active-area pitfalls. Both panels
schematic; energy levels not to scale.}
\label{fig:fund}
\end{figure}

Because the zero-bias signal is small and defect-sensitive, the metrics used
to report it are unusually easy to overstate. Several recurring pitfalls (the four most common summarized in Fig.~\ref{fig:fund}b and Table~\ref{tab:metrics}) deserve explicit attention.

\paragraph{Responsivity and irradiance.} The responsivity
\begin{equation}\label{eq:responsivity}
R=\frac{I_\mathrm{ph}}{P_\mathrm{opt}\,A},
\end{equation}
where $I_\mathrm{ph}$ is the photocurrent, $P_\mathrm{opt}$ the incident optical
power density (irradiance), and $A$ the active area, is typically sublinear in
intensity in trap-rich oxides, so a value at low
irradiance can exceed one at ``1~sun'' by orders of magnitude; the irradiance,
and ideally the calibration of the light source, must be
stated~\cite{wang2023characterize}.

\paragraph{Detectivity and noise.} Specific detectivity is frequently estimated
from the dark-current shot-noise term alone,
\begin{equation}\label{eq:detectivity}
D^*=\frac{R\sqrt{A}}{\sqrt{2qI_d}},
\end{equation}
omitting the $1/f$ and generation--recombination noise that dominate at low
frequency in defective oxides and inflating $D^*$ by one to two orders of
magnitude~\cite{wang2023characterize, dearquer2017solution}. Devices reporting a
measured noise spectrum or noise-equivalent power (NEP) (low-noise $\beta$-Ga$_2$O$_3$ arrays~\cite{tak2021noise} and ultralow-dark-current solar-blind diodes~\cite{sheoran2022zero}) give far more defensible figures, so we avoid
ranking on shot-noise-only $D^*$.

\paragraph{Speed and the pyroelectric artefact.} A polar layer produces a sharp
current spike synchronized with the light transient (the pyro-phototronic current) that is easily misread as an
intrinsically fast \emph{steady-state}
response~\cite{ma2017photovoltaic, jia2025pyro}. Transient and steady-state
contributions should be reported separately, the rise-time definition
(10--90\,\%, $1/e$, or RC-limited) stated, and setup-limited timing distinguished
from device-limited~\cite{jia2025pyro, jain2024heterostructured}.

\paragraph{Active area.} $R$ and $D^*$ scale with the area in the denominator, and
the device, illuminated, and electrode areas can differ by orders of magnitude;
the definition used must be stated~\cite{wang2023characterize}.

\paragraph{Linearity and frequency-domain reading.} A single responsivity number
presumes a linear response, yet trap-mediated oxides are frequently sublinear, so
the linear dynamic range (LDR) over which $R$ is constant should be reported
alongside it~\cite{tak2021noise}; stating the intensity dependence exposes the
sublinearity that inflates low-irradiance responsivities. Several pitfalls also
ease if the detector is read as a modulated device: chopping the light and
recovering the photocurrent at the modulation frequency rejects the $1/f$ and
drift terms that dominate at DC and separates the fast, field-driven component
from the slow trap-mediated tail, so the roll-off frequency doubles as a
device-limited speed check~\cite{kumar2021alternating}, provided the setup
bandwidth is stated.

\paragraph{External quantum efficiency and the gain question.} Reporting the
external quantum efficiency (EQE),
\begin{equation}\label{eq:eqe}
\mathrm{EQE}=\frac{hc}{q\lambda}\,R,
\end{equation}
alongside $R$ is the single cheapest guard against an implausible claim, because
it converts a responsivity into a photon-counting statement that a reader can
sanity-check: a junction device operating purely photovoltaically at zero bias
cannot exceed unity, so an $\mathrm{EQE}>1$ is a positive assertion that a gain
mechanism is present and obliges the authors to identify it. At zero bias the
admissible candidates are few, since the photoconductive recirculation that
supplies gain under bias is largely unavailable (Section~\ref{sec:fundamentals});
a transient pyroelectric contribution~\cite{ma2017photovoltaic, jia2025pyro}, a
photoelectrochemical process in which the electrolyte regenerates the surface
species~\cite{kaur2023review}, or trap-assisted barrier lowering at a contact are
the usual ones. Quoting $R$ alone leaves this unresolved; quoting EQE forces the
mechanism into the open, and a device whose EQE exceeds unity without a named
mechanism should be read as unexplained rather than as exceptional.

\paragraph{Operational stability.} The metric most often omitted is the one an
application depends on. A self-powered detector is proposed for deployments
lasting years, yet reported measurements are typically minutes of on/off cycling
under laboratory illumination, and the degradation risks noted in this review
(the ambient instability of halide perovskites, humidity-driven leakage in oxide
detectors) act on far longer timescales than that. Where devices are cycled extensively, the
resulting claim is correspondingly stronger: solar-blind Ga$_2$O$_3$ diodes characterized at elevated temperature (stable to
125~$^\circ$C)~\cite{sheoran2022zero}, and van der Waals photodiodes reporting stable
zero-bias output over repeated switching~\cite{song2021mxene}, are among the few
that can be compared on this axis at all. We therefore treat cycle count,
elapsed time, storage and operating atmosphere, and any encapsulation as part of
a performance claim rather than as supplementary information, and note where a
strategy's stability is asserted rather than measured.

These pitfalls are not unique to doped devices, but defect engineering amplifies
them: tuning dopants and vacancies changes dark current, noise, and transient
response at once, so a strategy can appear to improve $D^*$ or speed for reasons
that are artefacts rather than real gains. Taken together they also explain why
the quantitative cross-study compilation attempted in Section~\ref{sec:compare}
is harder than it appears: the pitfalls are not random scatter that averaging
removes, but systematic biases that inflate the best-looking numbers most, so a
ranking assembled from headline values is biased in a known direction. We flag
these distinctions throughout Section~\ref{sec:strategies}.

\begin{table}[htbp]
\centering
\small
\caption{Zero-bias metrics: what is commonly reported, the common pitfall,
and the correct practice adopted in this review.}
\label{tab:metrics}
\begin{tabular}{@{}p{2.3cm}p{4.3cm}p{4.6cm}@{}}
\toprule
Metric & Common pitfall at $V=0$ & Correct practice \\
\midrule
Responsivity $R$ & quoted at unstated/low irradiance; sublinear in power &
state irradiance and light-source calibration \\
Detectivity $D^*$ & shot-noise-only formula inflates $D^*$ &
report measured noise or NEP \\
Response speed & pyroelectric spike read as fast steady response &
separate transient/steady; define rise time \\
Active area & device vs.\ illuminated vs.\ electrode area conflated &
state the illuminated (active) area used \\
Quantum efficiency & $R$ quoted without EQE, hiding an implied gain $>1$ &
report EQE and name the gain mechanism \\
Linearity & single $R$ presumes linearity of a sublinear device &
report the linear dynamic range \\
Operational stability & minutes of cycling extrapolated to deployment &
state cycles, elapsed time, atmosphere, encapsulation \\
\bottomrule
\end{tabular}
\end{table}

\section{The family of engineering strategies}\label{sec:strategies}

We now treat each lever of Fig.~\ref{fig:umbrella} in turn. For every family we
state the operating principle, illustrate it with representative ZnO devices (drawing on wider-gap oxides as comparative benchmarks where ZnO examples are scarce, notably solar-blind Ga$_2$O$_3$), indicate the target metrics it serves
best, and note its characteristic limitations. The five families are not mutually
exclusive (the strongest devices usually combine two or three), but separating
them clarifies which
knob moves which metric, a mapping we draw together in
Section~\ref{sec:compare}.

\subsection{Junction and built-in field engineering}\label{sec:junction}

\begin{figure}[htbp]
\centering
\includegraphics[width=\linewidth]{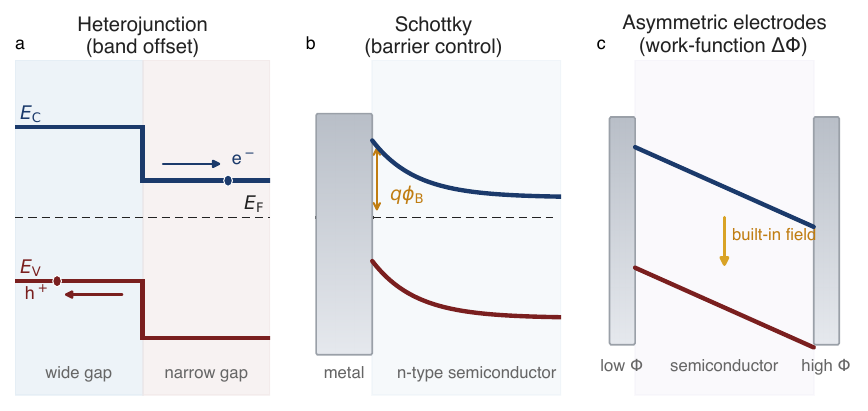}
\caption{Junction routes to a built-in field. (a) Type-II semiconductor
heterojunction band offsets separate electrons and holes; (b) a Schottky barrier of height
$q\phi_\mathrm{B}$; (c) two electrodes of different work function $\Phi$
producing a net built-in field across the semiconductor. Schematic.}
\label{fig:junction}
\end{figure}

The most direct way to create the required asymmetry is to engineer the junction
itself: choose materials and contacts whose equilibrium band alignment
establishes a built-in field across the absorber (Fig.~\ref{fig:junction}). This
is the default, most general, and usually most stable strategy, because the field
is set by thermodynamics rather than a metastable defect or polarization state.
Four sub-routes recur: \emph{(i) heterojunctions} pair two semiconductors with
staggered (type-II) or straddling (type-I) band edges so the offsets drive
electrons and holes apart (Fig.~\ref{fig:junction}a); \emph{(ii) Schottky
contacts} use a metal--semiconductor barrier tuned through the metal work function
and interface chemistry (Fig.~\ref{fig:junction}b)~\cite{liu2023review};
\emph{(iii) homojunctions and p--i--n stacks}: a homojunction places the field
inside a single material, avoiding lattice mismatch~\cite{zhang2025accelerating},
and a p--i--n stack spreads it across an intrinsic layer whose thickness can be
optimized~\cite{chen2023deep};
and \emph{(iv) asymmetric electrodes} exploit the work-function difference between
two dissimilar contacts to bias an otherwise symmetric film
(Fig.~\ref{fig:junction}c).

Band alignment is the primary variable: a type-II offset drives electrons and
holes into different layers, the alignment of choice for zero-bias separation,
though the effective optical gap becomes the difference between the two band edges,
red-shifting the onset and eroding solar-blind selectivity. In the Schottky route
the barrier height $q\phi_\mathrm{B}$ depends on the choice of metal and on
interfacial barrier modulation~\cite{liu2023review}, making
the interface engineering of Section~\ref{sec:interface} inseparable from junction
design.

Among heterojunctions, van der Waals (vdW) pairings have produced some of the most
efficient zero-bias UV devices: Ti$_3$C$_2$T$_x$ MXene/GaN and related 2D/mixed-dimensional photodiodes
deliver high efficiency with stable zero-bias output~\cite{song2021mxene, thota2023mxene}, and the same principle carries across the dimensional
combinations, with 2D MoS$_2$ on bulk i-GaN giving high responsivity at zero power
consumption~\cite{veeralingam2023responsivity}, a mixed-dimensional 1D~Se/2D~InSe
pairing doing likewise~\cite{shang2019mixed}, and GaN nanorods grown directly on a
transition-metal dichalcogenide inverting the usual stacking
order~\cite{zheng2022vertical}. Two features recur and are worth separating. The first is that a vdW interface tolerates lattice
mismatch~\cite{shang2019mixed}, so the pairing can be
chosen for its band offset rather than for its lattice constant, which is what
makes the design space so large. The second is that the 2D layer frequently does
double duty as a transparent charge-collection electrode, as in
graphene-contacted p-GaSe/n-InSe stacks with fast multicolour
response~\cite{yan2017multicolor}. For ZnO specifically, the vdW route is
accessible without a transfer step: MoSe$_2$/ZnO heterojunctions rectify and
detect over a broad band~\cite{wan2021mose2}, and an SnS$_2$/ZnO$_{1-x}$S$_x$
stack shows that alloying S into the ZnO side raises its band edges, reducing
interfacial recombination and shortening the rise time~\cite{jiang2020broadband}.

For solar-blind operation, wide-bandgap oxide heterojunctions dominate, and the
p-type partner is almost always a p-type oxide, because no wide-gap oxide can yet
be doped p-type reproducibly (Section~\ref{sec:doping}). All-oxide n-Ga$_2$O$_3$
paired with p-NiO is the reference couple~\cite{wang2020oxide}, and the same
p-NiO layer transfers to a spinel host in NiO/ZnGa$_2$O$_4$
diodes~\cite{khan2024unveiling}, illustrating how a single solved p-type contact
propagates through a whole family of n-type absorbers. Where a p-type oxide is
unavailable in film form, solution-processed p-type quantum dots supply the same
junction with far less thermal budget, as in p-MnO dots on
n-GaN~\cite{alamoudi2022solar} and on Sn-doped $\beta$-Ga$_2$O$_3$
microflakes~\cite{alwadai2023photoresponsivity}. A parallel line pairs the oxide
with a nitride instead: Ga$_2$O$_3$/AlN gives a high photo-to-dark ratio with fast
response~\cite{hao2021heterojunction}, and nanometre-thick amorphous
Ga$_2$O$_3$ on GaN shows that the absorber need not be thick when the field is
correctly placed~\cite{lin2023nanoscale}, a direct expression of the
absorption/collection overlap argument of Section~\ref{sec:fundamentals}. Sulfide
decoration of the oxide surface offers a further variant in which charge transfer
across a nano-interface, rather than a planar band offset, supplies the
asymmetry~\cite{xi2025nano}. Dual-junction designs can superpose two built-in fields to raise the zero-bias
responsivity~\cite{zhu2024ultrasensitive}, while a
same-material homojunction avoids lattice mismatch altogether~\cite{zhang2025accelerating}
and a PIN stack gives a larger built-in potential and space-charge region than a
Schottky contact on the same absorber~\cite{zhou2025mgzno}.

A recurring practical constraint deserves separate mention, because it is a
collection problem rather than a junction problem: in a wide-gap oxide the
generated carriers must still reach an electrode, and an opaque metal contact
placed over the illuminated face defeats the purpose. Interposing a transparent,
highly conductive charge-collection medium solves this without altering the
junction physics, and graphene is the common choice, whether as a collection layer
on an SnO$_2$ microwire/p-GaN device~\cite{xu2022sno2} or, nitrogen-doped and thus
p-type, as the junction partner itself against a $\beta$-Ga$_2$O$_3$
microwire~\cite{wang2022responsivity}. The distinction matters when reading such
reports: in the first case graphene is a contact and the junction is the
oxide/nitride interface, in the second the graphene \emph{is} the p-side, and the
two have different stability and reproducibility profiles. Finally, once several
junctions of different gap can be integrated on one chip, the spectral axis becomes
addressable rather than fixed, and dual-sided Ga$_2$O$_3$/GaN p--n arrays have been
operated as a multiband UV communication link~\cite{shi2025dual}, an application
where zero-bias operation and wavelength discrimination are both requirements
rather than conveniences.

These examples form a design gradient: at one end the vdW heterojunctions, chosen for band
offset rather than lattice constant, at the other the all-oxide couples, which accept a more defective interface for
robustness, solar-blind gaps, and vapour- or solution-compatible growth. A
separate design axis is spectral selectivity: coupling two junctions of different gap lets the
sign or magnitude of the zero-bias photocurrent encode
wavelength~\cite{he2019nanorod}, as does a polarity-flipping
absorber whose field reversal gives wavelength-selective response from a single
junction~\cite{goswami2024polarity}, well suited to the solar-blind
and multiband targets of Section~\ref{sec:compare}.

At the low-cost end sit the Schottky and asymmetric-electrode routes, which impose
the field with a single deposited contact rather than a matched second
semiconductor: a Schottky barrier needs only one rectifying metal, while an
asymmetric-electrode device biases a symmetric film purely through the
work-function difference between two dissimilar contacts. They trade peak built-in
potential for fabrication simplicity (fewer deposition and alignment steps than
any junction pairing), which is why they recur in low-cost and doped-oxide
detectors~\cite{liu2023review, tsay2022improved}; their weakness is the
thin depletion layer and low barrier noted above~\cite{liu2023review}, which tie
the delivered field to interface preparation and hence to reproducibility. They combine naturally with the
doping and architecture families when manufacturability, not maximum
responsivity, is the goal.

Junction engineering is the strategy of choice when steady-state responsivity and
long-term operational stability are the priorities, and it imposes the fewest material constraints. Its
limitations are equally clear: the built-in voltage, and hence the maximum
zero-bias collection, is capped by the available band offset or barrier height;
type-II alignment that improves separation also narrows the optical gap; and
lattice or thermal mismatch at the interface seeds the defect states that dominate
dark current and recombination~\cite{jia2025engineering}. Hybrid designs that add
a second field (a ferroelectric layer for tunability~\cite{chen2020highly}, or a sharpened offset across a Ga$_2$O$_3$/GaN heterointerface~\cite{varshney2022heterointerface}) recover some headroom.
These interface defects bridge to the next two families, which manage them
deliberately.

\subsection{Doping and defect engineering}\label{sec:doping}

Doping is the most accessible lever, particularly for the solution-processing
laboratories that produce much of the oxide-detector literature, and the
zero-bias doping paradox of Section~\ref{sec:paradox} makes it the most subtle.
Rather than catalogue dopants by element, we organize them by the function they
perform in the device (Fig.~\ref{fig:doping}b).

\paragraph{Carrier-density tuning.} Shallow donors (Al, Ga, In in the case of
ZnO) raise the free-electron density and the built-in field but, through $W\propto N_d^{-1/2}$,
narrow the collection region, the volcano dependence of
Section~\ref{sec:paradox}, so the response peaks at an intermediate concentration.
AZO-seeded ZnO nanorod arrays report such a window, 0.5\,at.\% Al outperforming
both the undoped and the 1\,at.\% seed~\cite{zhang2016development}, as does a
Bi-doping series of SnO$_2$ that performs best at 4\,mol\%~\cite{huang2025boosting}. Heavy doping also degrades transport:
ionized-impurity scattering lowers mobility and Burstein--Moss filling blue-shifts
the onset, so electrical and optical penalties compound the geometric one, the
clearest case of ``less is more'' at zero bias. The author's spray-deposited
Al:ZnO films sit in this regime, their response set by doping-tuned carrier
density and crystallinity together~\cite{nurfani2020al}.

\paragraph{Band-edge engineering.} Alloying and doping shift the absorption
edge, giving the spectral selectivity that solar-blind detection demands. Alloying ZnO with Mg widens its
gap, and at high Mg content the cubic MgZnO alloy reaches the far-UVC in Schottky
and PIN devices~\cite{zhou2025mgzno}; co-sputtered Y--In oxide alloys tune the gap
to 4.70~eV for the solar-blind window~\cite{gan2026bandgap}; and Zr doping combined
with an organic partner gives a dual-band UV--visible response~\cite{li2025visible}.
Doping does not always move the edge monotonically: in the author's Mg-doped ZnO
nanorods the gap widens at 1\% Mg but narrows again at 3--5\% as the rods grow,
while the UV sensitivity keeps rising~\cite{nurfani2024mg}. Band-edge alloying
also changes the host phase at high content~\cite{zhou2025mgzno}, which bounds how
far the edge can be pushed in a given crystal structure.

A subtlety is worth separating from the gap shift itself. Within the alloy range
the dopant does not simply translate the absorption edge: it also reshapes
the excitonic structure that sits at the edge, and in a wide-gap oxide with a
large exciton binding energy this determines how sharp the spectral cut-on
actually is. Isovalent and higher-valent substitutions in ZnO alter excitonic
oscillator strength and screening rather than only the gap~\cite{nurfani2019znosn, nurfani2018screening}, so two films with the same nominal
optical gap can differ in the steepness of their band edge, and therefore in the
UV/visible rejection ratio that defines a selective detector. For solar-blind
targets, where the specification is a rejection ratio and not merely a cut-on
wavelength, the edge sharpness is the operative quantity. Dopant levels can equally move the
response away from the UV: indium doping of amorphous boron nitride creates defect
transitions that enable a self-biased, fast 700--1000~nm detector on
silicon~\cite{li2025multicolor}.

\paragraph{Deep-level and defect modulation.} Transition-metal dopants (Cu, Fe,
Ni, Co, Mn) and native point defects introduce mid-gap states with opposite
effects on the two metrics that matter: a level near a band edge is a shallow trap
that stores and re-emits one carrier type, lengthening the response tail, whereas
a mid-gap level with comparable electron and hole cross-sections is an efficient
Shockley--Read--Hall centre that removes carriers before collection, lowering
responsivity but often quickening the device by suppressing the slow tail. In
oxides this is entangled with persistent photoconductivity, where photodesorbed
surface oxygen~\cite{boruah2019zinc} or oxygen vacancies~\cite{chen2025spiro}
leave carriers that decay slowly; because carriers trapped in deep sites slow the
transient~\cite{ocaya2023bias}, the deep-level population sets zero-bias speed as
directly as the junction does. Independent studies of transition-metal- and
defect-engineered ZnO photodetectors report the same dual role of mid-gap
states~\cite{nguyen2021visible, muhammad2021fluorine, kumar2024defect}. This
sub-family is the author's main line of work, and transition-metal-doped ZnO, mostly grown
by spray pyrolysis (Ti by sputtering), spans the whole class. Copper
is the clearest dual case: a 0--2.5\,at.\% series tunes both the
Schottky-barrier height~\cite{nurfani2020cu} and the UV
sensitivity~\cite{nurfani2021cu}, while Fe, Ti, and Mn modulate the same
properties~\cite{nurfani2021fe, nurfani2017ti, pertiwi2024interlayer} (Ti reshaping the excitonic and defect states directly, with defect-induced recombination channels resolved spectroscopically in Ti$_x$Zn$_{1-x}$O~\cite{nurfani2018screening, nurfani2017tidefect}), culminating in self-powered Mn- and Ni-doped ZnO
heterojunctions~\cite{pertiwi2024interlayer, nurfani2025nizno}. The numbers make
the trade-offs concrete: UV sensitivity peaks at 1.5\,at.\% Cu within a
0--2.5\,at.\% window~\cite{nurfani2021cu}, Fe raises it up to seventyfold at
3\,at.\%~\cite{nurfani2021fe}, and 3\%\,Ti suppresses the dark current and lifts
the photo-to-dark ratio~\cite{nurfani2017ti}; the same dopants reshape the grains
(Cu ${\sim}114\to156$~nm, Fe ${\sim}67$~nm, Ni roughly halving the
nanostructure)~\cite{nurfani2021cu, nurfani2021fe, nurfani2025nizno}, shifting the
grain-boundary trap reservoir with the intragrain deep levels. The
responsivity--speed tension then appears explicitly at zero bias: in Ni-doped
ZnO/p-Si the undoped film gives the highest photo-to-dark ratio (162 versus 16 at
3\%\,Ni) yet the doped device is over an order of magnitude faster (0.4 versus
10.8~s rise)~\cite{nurfani2025nizno}, while in sputtered films oxygen-vacancy
states underlie a ${\sim}5$~ms response~\cite{nurfani2018fast}. Reproducibility is
the standing caveat: deep-level concentrations depend on batch, atmosphere, and
thermal history, magnetic dopants especially.

Read constructively, this dual role becomes a design tool rather than a defect to
be minimized. If a trap distribution can be made deliberately asymmetric between
electrons and holes, the resulting imbalance in transport is itself a
selectivity mechanism: an unbalanced-carrier-transport design has been used to
turn a broadband absorber into a narrowband red/UV detector without any optical
filter~\cite{hou2020narrowband}. The same principle underlies the responsivity--speed
choice above, where a mid-gap centre is accepted in exchange for a shorter tail.
What distinguishes a designed defect landscape from an accidental one is
therefore not defect density but \emph{control}, and the practical difficulty is
that the characterization techniques able to demonstrate such control
(Section~\ref{sec:designrules}) are rarely applied.

\paragraph{Native defects and growth-side control.} Because dopant and
native-defect populations are co-determined by growth, the cheapest defect knob is
often the deposition itself. In spray-pyrolysed ZnO the substrate temperature
alone governs the defect density, electronic transitions, and (crucially for zero-bias detectivity) the dark current~\cite{nurfani2019substrate}, while
surface modification and growth arrangement tune the oxygen-vacancy population
behind the green-emission and trap landscape of ZnO films and
nanorods~\cite{nurfani2018ovac, purbayanto2020greenlum, purbayanto2019nanorods};
the origin of the fast photocurrent has been
traced in the same films~\cite{nurfani2018fast}.

Four growth variables recur, and each maps onto a distinct part of the defect
landscape. \emph{Substrate temperature} sets the adatom mobility during
deposition and hence the vacancy population and dark current~\cite{nurfani2019substrate}. \emph{Precursor molarity} sets the arrival rate and
thus grain size and porosity, and in sprayed ZnO it measurably shifts the optical
and photoactive response~\cite{nurfani2023molarity}. \emph{Post-deposition
annealing} redistributes the defects already present rather than changing their
number wholesale, which is why it is such an effective late-stage correction: a
simple post-anneal raises the UV sensitivity of ZnO:Cu~\cite{nurfani2022cu}, and
in Ti-doped ZnO it measurably reorganizes the excitonic and defect
states~\cite{darma2019postann}, while in hydrothermally grown ZnO the combination
of solution pH and post-annealing controls the resulting crystal habit~\cite{oktapia2023seedless}. \emph{Morphology and orientation} close the set:
a seed layer reshapes sprayed ZnO:Fe and raises its UV
sensitivity~\cite{nurfani2021feseed}, hydrothermal pH tunes the nanorod structure
that governs UV detection~\cite{nurfani2025phzno}, and the in-plane arrangement of
nanorods alters the optoelectronic response through the point-defect population it
carries~\cite{purbayanto2019nanorods}.

The practical significance is that these four knobs are available in any
solution-processing laboratory~\cite{dearquer2017solution}, require no additional
precursor, and act on the
same deep levels the dopants target~\cite{nurfani2019substrate}. They are also the
knobs least often reported
in enough detail to be reproduced, which is why the batch-to-batch variation
noted in Section~\ref{sec:designrules} is so persistent: a paper that specifies
its dopant concentration to two decimal places but not its substrate temperature,
molarity, ramp rate, or annealing atmosphere has specified the smaller half of its
defect chemistry. Substrate temperature, molarity, morphology, and annealing thus
offer a low-cost, dopant-free route to the same defect landscape, the natural
companion to the dopant toolbox in solution-processing laboratories.

\paragraph{Doping as a modifier of the other families.} Doping is rarely deployed
alone, and its most effective uses in the recent literature are as a modifier
applied \emph{inside} another strategy, which is also why it resists being ranked
against them. Within the photoelectrochemical family, cation doping of the
photoanode raises the zero-bias response of $\alpha$-Ga$_2$O$_3$ nanowire and
nanorod arrays, whether by aluminium during hydrothermal
growth~\cite{guo2023hydrothermal} or by indium~\cite{xue2025boosting}, and silver
doping activates a ZnO/MoS$_2$ nanoarray photoelectrode~\cite{kuang2025doping}.
Within the architecture family, doping tunes the conductivity of the collecting
network rather than the absorber, as with boron and nitrogen co-doped reduced
graphene oxide~\cite{yilmaz2025simultaneous} or \emph{in situ} salt-doped
conducting-polymer nanowire networks formed during spin coating~\cite{kim2026situ}.
The pattern is consistent: doping supplies a tuning parameter to whichever field
source has already been chosen, and the zero-bias paradox constrains it in every
case, so it is best read as a modifier axis crossing Fig.~\ref{fig:umbrella}
rather than as a competing alternative to the other four families.

\paragraph{Polarization enhancement.} Dopants that strengthen the wurtzite
polarization or modify the pyro/piezo coefficients couple doping to the
polarization family of Section~\ref{sec:polarization}; a systematic study of
doping concentration on the pyro-phototronic response makes this link
quantitative~\cite{wan2021doping}, as does V-doped ZnO, whose own
ferroelectricity and piezoelectricity add to the built-in field of a Ga$_2$O$_3$/ZnO:V
junction~\cite{wang2024heterojunction}.

\paragraph{p-type doping and the homojunction problem.} Robust p-type doping of
ZnO and Ga$_2$O$_3$ remains the central unsolved problem, and it is fundamental:
acceptor levels lie deep above a low-lying valence band and ionize poorly, while
native donors (oxygen vacancies, metal interstitials) form readily and
self-compensate any acceptors, so hole doping is thermodynamically costly and
rarely stable. Nitrogen and co-doping, and p-type partner oxides (NiO, CuO,
Cu$_2$O), are the practical routes~\cite{chen2025spiro, xu2024spraying}; copper iodide
(CuI), an intrinsically p-type transparent semiconductor, illustrates the payoff
when a stable p-type partner is available, a sprayed transparent Sn-doped CuI
film driving an ultrahigh-speed self-powered detector~\cite{xu2024spraying}. The
prize for solving it is visible in the rare cases where a ZnO homojunction has
been realized: low-temperature-grown p--n ZnO nanojunction arrays operate as
rapid, self-driven UV detectors~\cite{lu2016temperature}, combining a
lattice-matched interface, a single thermal budget, and no second material to
mismatch or delaminate. That such devices remain isolated demonstrations rather
than a device family is the clearest measure of how far from routine the
underlying doping problem is. Until reproducible p-type wide-gap oxides exist,
heterojunction partners remain the default.

Two subtleties recur. The useful dopant is often a \emph{combination}: co-doping
decouples roles one dopant conflates (a shallow donor raising the field while a
second element compensates a deep level or relieves strain), which is why Ga--In
and Al--Cu co-doped ZnO recur~\cite{tsay2022improved, tsay2020improving, demirbilek2021structural}. And in polycrystalline, solution-grown oxides, grain
boundaries form a trap reservoir that co-determines the deep-level landscape, so
zero-bias speed and dark current depend on grain size and boundary chemistry as
much as on nominal doping~\cite{nurfani2019substrate, nurfani2018ovac}. Doping is
therefore best understood not as a single species but as a dopant--defect ensemble
whose optimum is zero-bias-specific: the cheapest route to spectral selectivity
and response-time control, but carrying the paradox, the reproducibility burden,
and the p-type bottleneck as standing costs.

\subsection{Polarization coupling}\label{sec:polarization}

\begin{figure}[htbp]
\centering
\includegraphics[width=\linewidth]{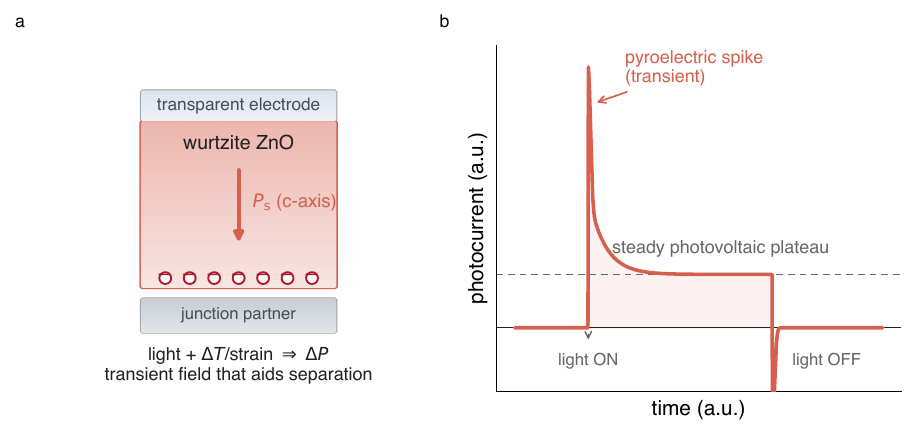}
\caption{Polarization coupling. (a) Spontaneous polarization $P_\mathrm{s}$ in
a wurtzite layer adds bound interface charge that steepens the built-in field.
(b) The pyro-phototronic current appears as a sharp transient spike at each
light transition, superposed on the steady photovoltaic plateau; the two must be
reported separately. Schematic; arbitrary units.}
\label{fig:polar}
\end{figure}

A polar layer (pyroelectric, piezoelectric, or ferroelectric) contributes a
bound-charge field that can augment or even replace the junction field
(Fig.~\ref{fig:polar}a). The pyro-phototronic effect converts the transient
temperature change on illumination into a polarization-charge pulse that sharpens
carrier separation, producing the characteristic current spike at each light
transition~\cite{ma2017photovoltaic, jia2025pyro}. In wurtzite ZnO and 2D
halide-perovskite ferroelectrics this transient can dominate the zero-bias output:
ZnO nanofibre arrays on BaTiO$_3$ exploit the effect
directly~\cite{yu2023pyro}, ferro-pyro-phototronic 2D perovskite microbelts and
films report large self-powered responsivities~\cite{guo2022ferro,
guo2023ultrahigh}, and pyroelectric coupling drives flexible broadband stacks on
Si/CdS~\cite{dai2018flexible}; beyond the transient, the ferroelectricity and
piezoelectricity of V-doped ZnO strengthen the field of a solar-blind
Ga$_2$O$_3$/ZnO:V junction~\cite{wang2024heterojunction}.

For a ZnO-based device the important point is that the polar term need not be
imported from a separate ferroelectric layer at all: wurtzite ZnO is itself
polar, and its net polarization depends on how the film is grown. Highly oriented
ZnO on Si shows enhanced polarity when the growth is controlled to align the
$c$-axis~\cite{kurniawan2016polarity}, and the ferroelectric-like response of
sputtered ZnO films is itself modified under
illumination~\cite{kurniawan2017ferro}, which is precisely the coupling this
family exploits. Two consequences follow for design. First, orientation control
belongs in the same toolbox as doping and annealing, since a textured film
supplies a bound-charge contribution that a randomly oriented one does not, at no
extra process cost. Second, the light-dependence of the polarization means the
polar term is not a fixed offset added to the junction field but is itself
modulated by the illumination being measured, so the transient and steady
responses are coupled rather than merely superposed, which is the physical
reason they are so easily conflated in reporting.

The recent literature has moved from demonstrating the effect toward using it as
a controllable design variable, and three directions are visible. The first is
\emph{integration with a conventional junction}, where the polar layer boosts an
existing field rather than supplying it: BaTiO$_3$@MXene on 4H-SiC, a UV detector~\cite{sha2024ferro}, and amorphous GaO$_x$
on a ferroelectric Hf$_{0.5}$Zr$_{0.5}$O$_2$ layer, a solar-blind
one~\cite{xu2025enabling}, both couple the ferro-pyro-phototronic effect to an
existing junction, the latter showing
that the poled state can invert the sign of the photoresponse, turning
polarization into a programmable rather than fixed parameter. The second is
\emph{polarization-sensitive detection}, in which the ferroelectric axis makes the
device respond to the polarization state of the incident light: 2D halide
perovskites give strongly polarization-sensitive zero-bias
photodetection~\cite{ji2020ferroelectricity}, and heterojunction designs achieve a controllable polarization
ratio~\cite{yang2025perovskite}, a functionality no other family in
Fig.~\ref{fig:umbrella} provides. The third is \emph{spectral extension}, where
the pyroelectric term, being thermal in origin, responds far beyond the absorption
edge and so broadens the detected range from the ultraviolet through the
infrared~\cite{tang2023photoexcited, fu2024visible}, while in inorganic
ferroelectrics a broad range is more often reached by pairing the polar layer with
a narrower-gap absorber, as in van der Waals LiNbO$_3$/WS$_2$~\cite{liang2023waals},
or by narrowing the ferroelectric's own gap, as in non-stoichiometric BZT--BCT
ceramics~\cite{wang2022stoichiometric}. For a UV-selective detector this breadth is a liability
rather than an asset, and it is worth stating plainly: a strong pyroelectric
contribution buys transient gain at the cost of spectral selectivity, because a
thermal response does not discriminate by photon energy.

The three effects differ in what drives the bound charge. The pyroelectric
contribution responds to the \emph{rate} $dT/dt$ as light switches, appearing as
opposite-sign spikes at the rising and falling edges and vanishing under steady
illumination, the four-lobe signature that must not be mistaken for a fast
steady-state response. The piezoelectric contribution responds to strain and gates
the band profile at a contact; the ferroelectric contribution adds a
\emph{switchable, remnant} polarization that biases the junction even in the dark,
so poling sets the sign and magnitude of the response. In 2D perovskite single
crystals the three coexist and their coupling multiplies the transient gain, a
progression from demonstrating the effect to polarization-\emph{sensitive}
detection that the oxide ferroelectrics are now following into the solar-blind
window.

Polarization coupling is the natural choice for pulsed or modulated detection (optical communication, chopped-source imaging), where the
transient spike is an asset, and for boosting an otherwise weak junction. It is also the family most
prone to the reporting pitfalls of Section~\ref{sec:metrics}: the
pyroelectric spike is a transient, not a steady-state, response
(Fig.~\ref{fig:polar}b), and conflating the two overstates speed and
responsivity~\cite{ma2017photovoltaic, jia2025pyro}. A further limit is lifetime: the halide-perovskite
ferroelectrics that give the largest responses remain hindered by long-term
instability under ambient conditions~\cite{veeramalai2021lead}.

\subsection{Interface, surface, and photoelectrochemical engineering}\label{sec:interface}

Where the previous families build a field in the bulk, this one engineers the
plane where separation occurs. Three routes are common. Surface \emph{passivation
and interlayers} neutralize mid-gap surface trap states: rinsing CsPbBr$_3$
nanocrystal films in excess bromide passivates them, suppresses hysteresis, and
raises the responsivity of a ZnO/CsPbBr$_3$ heterojunction
detector~\cite{saleem2023suppression}. \emph{Plasmonic}
nanostructures raise useful absorption without widening the semiconductor:
Ag@SiO$_2$ core--shell particles in a TiO$_2$ photoanode enhance light harvesting
and inhibit recombination~\cite{huang2017plasmon}, an approach that transfers directly to nanostructured oxide
detectors, as in Ag-nanoparticle-decorated Gd$_2$O$_3$
nanorods~\cite{meitei2023nanoparticle}. Most distinctively,
\emph{photoelectrochemical} detectors replace the solid junction with a
solid/liquid interface: equilibration of the oxide Fermi level with the
electrolyte redox potential bends the bands and forms a depletion layer while the
Helmholtz double layer accommodates the drop on the liquid side; for the n-type
oxide photoanodes considered here, holes drive a surface oxidation while electrons
exit through the back contact, so separation occurs across a step that costs
almost nothing to fabricate and is intrinsically large for
wide-gap oxides~\cite{zhou2015overview, kaur2023review, chen2019photoelectrochemical}.

The PEC route has developed along three axes worth separating, because they
impose different constraints. The first is the \emph{photoanode material}, where
the requirement is a wide gap combined with chemical durability in the
electrolyte: $\alpha$-Ga$_2$O$_3$ nanorod photoanodes, including wire-shaped
geometries that dispense with a planar substrate
entirely~\cite{wang2023wire}, In$_2$O$_3$ nanocube films~\cite{zhang2024photoelectrochemical}, and ultrathin Bi$_2$O$_2$S
nanosheets for broadband operation~\cite{yang2022broadband} span the range from
solar-blind to broadband, while ZnO itself serves as the nanostructured scaffold
in sensitized architectures~\cite{altaf2024photoelectrochemical}. Surface
chemistry matters as much as the bulk: suppressing interfacial doping raises the
photoresponse of InSe photoelectrodes~\cite{yang2022boosting}, and a
methyl-terminated germanene surface resists the degradation that limits less
stable photoanodes~\cite{xi2025methyl}. The second axis is the
\emph{electrolyte}, the component with no analogue in a solid-state device.
Because it is a designed part of the junction rather than a passive medium, it can
be chosen for the deployment: seawater has been used directly as the electrolyte
for portable and wireless optical links~\cite{luo2022demonstration}, which turns
the packaging problem of an underwater sensor into an advantage, and
(In,Ga)N-nanowire photoelectrodes have been operated for under-seawater
detection~\cite{zhang2023nanowire}. The third axis is \emph{speed}, historically
the PEC family's weakness, since the response involves ionic transport and
interfacial reaction rather than carrier drift alone; recent GaN photoelectrodes
address this trade-off explicitly, breaking the responsivity--bandwidth limit
sufficiently for optical communication~\cite{fang2023breaking}, and vertically
aligned GaS nanosheets combine fast response with underwater
operation~\cite{liu2024vertically}.

PEC devices are the cheapest path to solar-blind detection and suit liquid-phase
and underwater optical communication, and hydrogel-electrolyte variants make
them suitable for wearable applications~\cite{li2024flexible, chen2025quasi}. Their limits are chemical,
not electronic: the electrolyte is best viewed as a tunable part of the junction,
whose kind and concentration change both the photoresponse and its
stability~\cite{he2024influence, zhang2025optimizing}, and a too-aggressive
electrolyte can corrode the photoelectrode, as ZnS does in concentrated
H$_2$SO$_4$~\cite{zhang2025optimizing}. This chemical sensitivity can also be turned into a design
freedom: a quasi-solid hydrogel adds flexibility and amphibious operation
without losing the photoresponse~\cite{li2024flexible, chen2025quasi}, while an $\alpha$-Ga$_2$O$_3$ nanorod
photoanode on a flexible Ti substrate stays stable through repeated bending and
switching~\cite{chen2023flexible}. For these surface-dominated devices, defect and
interface control is the primary design variable, not an enhancement.

\subsection{Architecture and light management}\label{sec:architecture}

The final family is geometric: shaping the absorber and contacts to harvest more
light, collect carriers over shorter distances, and enable flexible, large-area,
low-cost manufacturing. Nanostructured and core--shell morphologies increase
junction area and shorten the carrier path (ZnO/ZnS~\cite{lin2016core} and
ZnO--Co$_3$O$_4$~\cite{ghamgosar2019core} core--shell arrays and metal-oxide
nanobrushes~\cite{jain2024heterostructured} raise the zero-bias response relative
to planar films), while single-nanowire and microwire detectors maximize the field
per unit volume in a minimal footprint~\cite{ozdemir2021responsivity, yildirim2021fine}.

The core--shell geometry deserves separate comment, because it is the clearest
structural expression of the collection argument of
Section~\ref{sec:fundamentals}: the junction is wrapped around the absorber, so
every generated carrier is within a short radial distance of the field regardless
of where along the wire it was created. Which shell is chosen then sets what the
device is for. A wider-gap shell on a ZnO core passivates and adds a
type-II offset~\cite{lin2016core}; a p-type oxide shell such as Co$_3$O$_4$ turns
the same array into an all-oxide p--n junction~\cite{ghamgosar2019core}; a
narrower-gap shell such as CdS extends the response to a second wavelength band,
giving dual-wavelength discrimination from one structure~\cite{sarkar2015highly};
and a sub-stoichiometric oxide shell such as MoO$_{3-x}$ on GaN supplies a high
work function that deepens the band bending~\cite{zheng2020core}. The same
principle carries beyond the oxides in Si/Cu(In,Ga)Se$_2$ core--shell broadband
detectors~\cite{chen2019cuin}. Porosity achieves a related effect in planar
films by lengthening the optical path without lengthening the collection path, and
multiple-stacked nano-porous ZnO shows the associated enhancement in emission and
excitonic response~\cite{darma2018porous}.

Two mechanisms underlie these gains, both rooted in the zero-bias collection
physics of Section~\ref{sec:fundamentals}. Nanostructuring \emph{orthogonalizes}
absorption and collection: a nanorod absorbs along its length while carriers are
collected across its much shorter radius, so strong absorption coexists with a
short drift distance, the one escape from the doping paradox open to a planar
film, since it enlarges the effective collection volume without raising $N_d$;
core--shell geometries extend this by wrapping the junction around the
absorber~\cite{lin2016core}. The cost is
surface: the same high surface-to-volume ratio multiplies the states that seed
recombination and persistent photoconductivity, so the fastest nanostructured
devices are also the most heavily passivated~\cite{jain2024heterostructured},
tying this family back to the surface engineering of Section~\ref{sec:interface}.

Beyond raising single-device metrics, architecture also sets the form factor, and
this family is indispensable when flexibility, wearability, or large-area coverage
is the target: flexible detectors on carbon cloth and on titanium foil demonstrate robust
self-powered operation~\cite{sun2017flexible, chen2023flexible}, and a p--n detector
that is self-powered in planar form has been turned into a fibre-shaped, real-time
wearable UV monitor~\cite{xu2018real}.

Form factor is not a single requirement but several, and the reported
architectures separate accordingly. Where the need is \emph{conformability},
one-dimensional nanostructures on compliant substrates are the established
route~\cite{lou2015flexible}, whether as a single patterned
nanowire~\cite{yildirim2021fine}, quantum-dot Schottky heterojunctions on a
flexible foil~\cite{praveen2021flexible}, or nanowires released from their growth
substrate and re-embedded in a hydrogel matrix, which reduces device volume while
accelerating the response~\cite{jiang2022flexible}. Where the need is
\emph{textile integration}, the detector must survive weaving rather than
bending, and fibre-based devices meet this directly, both as woven fibrous arrays
for UV optical communication~\cite{deng2023woven} and as carbon-nanotube woven
textile photodetectors~\cite{zubair2018carbon}; biocompatible hybrids such as
gold-nanoparticle-embedded silk/ZnO-nanorod structures extend the same idea toward
skin-mounted use~\cite{gogurla2017gold}. Where the need is \emph{transparency or
omnidirectionality}, the constraint shifts to the electrodes and the substrate,
and printable transparent detectors with omnidirectional response~\cite{lien2018omnidirectional} and transparent conducting-oxide-based flexible
stacks~\cite{xu2020perovskite} address it. Where the need is \emph{large-area
manufacture}, full printing is the endpoint, and fully printed flexible
solar-blind detectors built from atomically thin wide-gap
heterostructures~\cite{lu2025fully} show that the wide-gap and the printable
requirements are not mutually exclusive. Hybrid schemes such as hydrothermally
processed nitrogen-doped graphene quantum dots on silicon nanowire arrays pair a
solution-processed UV absorber and down-converter with a nanostructured silicon
collector~\cite{mondal2021silicon}.

Electrospinning deserves mention as a scalable route to precisely the
high-surface-area, fibre-based morphology this family exploits: it deposits
continuous nanofibre mats over large areas from solution at low temperature and
is already established for functional sensing layers~\cite{rianjanu2021electrospun}, and the mechanical robustness of the resulting
mats, the usual objection to their use in a working device, can itself be
engineered through post-spinning solvent treatment~\cite{rianjanu2018solvent}.
Applied to oxides, it produces the ZnO nanofibre morphology used in the
polarization-coupled detectors of Section~\ref{sec:polarization}~\cite{yu2023pyro}.
The broader point is that the solution chemistry which builds oxide/oxide
heterojunctions for other photoactive applications transfers directly to detector
stacks, since both depend on interfacial charge separation across the same
junction: hydrothermally grown CeO$_2$/ZnO heterojunctions, developed for
photocatalytic use~\cite{khan2025ceozno}, are built by the same route and at the
same thermal budget as the detector architectures above. It is where the author's solution-processing
perspective is most natural, and where the sustainability case is sharpest: a
battery-free device that still needs molecular-beam epitaxy or a transferred 2D
flake has relocated the cost, not removed it, whereas spray pyrolysis,
chemical-bath and hydrothermal growth, and full solution printing deposit doped,
nanostructured oxides over large flexible areas at low temperature and low capital
cost~\cite{dearquer2017solution, jiang2024solution}, sprayed doped-ZnO films and
hydrothermally grown ZnO nanorod arrays being canonical
examples~\cite{nurfani2020al, nurfani2025phzno}. The penalty, defect density and batch variation, returns to doping and interface control:
the cheap, green route and the defect-engineering burden are two faces of the same
choice, and the reproducibility tension is what the design rules of
Section~\ref{sec:designrules} must resolve.

\section{Comparing strategies across target metrics}\label{sec:compare}

\begin{figure}[htbp]
\centering
\includegraphics[width=\linewidth]{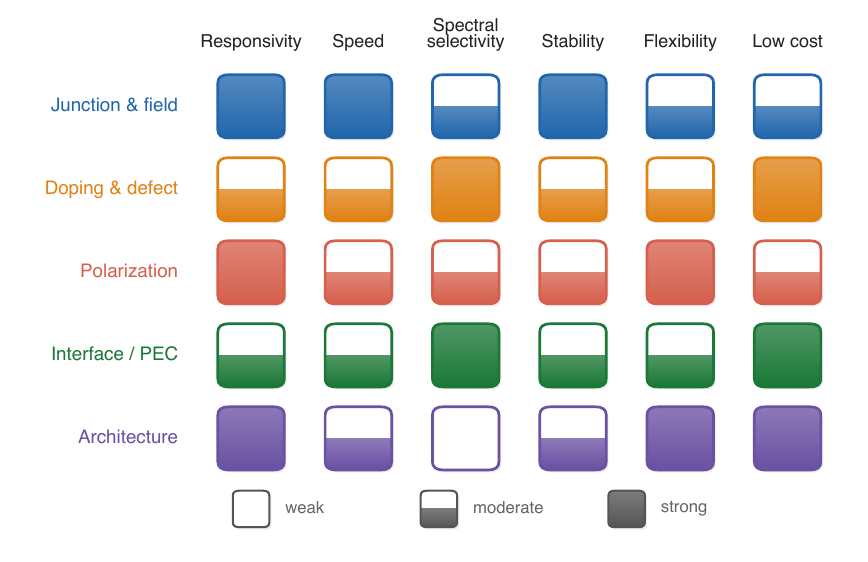}
\caption{Qualitative strategy--metric matrix. Each cell is a
gauge whose fill level (empty/half/full) encodes how strongly a strategy family
tends to move a target metric (weak/moderate/strong), synthesized from the device
literature of Section~\ref{sec:strategies}. It is a design heuristic, not measured
data.}
\label{fig:matrix}
\end{figure}

No single family wins on every axis; the practical question is which lever to pull
for a given target, and what it costs elsewhere. Table~\ref{tab:families}
first summarizes the five families along the axes a designer commits to in
sequence: what physically supplies the built-in asymmetry, which target metric the
family serves best, what it costs, and which growth route it presumes.
Figure~\ref{fig:matrix} then
synthesizes Section~\ref{sec:strategies} into a strategy--metric matrix, a
design heuristic synthesized from the representative literature, not a statistical
average, most usefully read one target metric (column) at a time.

\begin{table}[htbp]
\centering
\small
\caption{The five engineering-strategy families compared along the axes a
designer commits to. Entries summarize the mechanisms and trade-offs established
in Section~\ref{sec:strategies}; they are qualitative design guidance synthesized
from the cited literature, not measured performance data.}
\label{tab:families}
\begin{tabular}{@{}p{2.1cm}p{2.7cm}p{2.4cm}p{2.6cm}p{2.9cm}@{}}
\toprule
Family & Source of asymmetry & Serves best & Chief limitation &
Representative systems \\
\midrule
Junction \& built-in field (\S\ref{sec:junction}) &
equilibrium band alignment: p--n, Schottky, type-II, asymmetric contacts &
steady-state responsivity; operational stability &
$V_\mathrm{bi}$ capped by available offset; interface defects &
NiO/Ga$_2$O$_3$~\cite{wang2020oxide}; MXene/GaN~\cite{song2021mxene};
MoSe$_2$/ZnO~\cite{wan2021mose2} \\
\addlinespace
Doping \& defect (\S\ref{sec:doping}) &
modifies $N_d$, band edge, and deep levels of an existing field &
spectral selectivity; response-time control &
zero-bias paradox; reproducibility; no p-type oxide &
Cu-, Fe-, Mg-doped ZnO~\cite{nurfani2021cu, nurfani2021fe, nurfani2024mg};
Ga--In:ZnO~\cite{tsay2022improved} \\
\addlinespace
Polarization (\S\ref{sec:polarization}) &
bound charge: pyro-, piezo-, ferroelectric &
pulsed and modulated detection; polarization sensitivity &
transient/steady conflation; thermal response is not
UV-selective &
ZnO/BaTiO$_3$~\cite{yu2023pyro}; GaO$_x$/HZO~\cite{xu2025enabling};
2D perovskites~\cite{guo2022ferro} \\
\addlinespace
Interface \& PEC (\S\ref{sec:interface}) &
surface states removed; solid/liquid potential step &
low-cost solar-blind; underwater and liquid-phase use &
electrolyte stability and packaging; ionic speed limit &
$\alpha$-Ga$_2$O$_3$ photoanodes~\cite{wang2023wire};
seawater electrolyte~\cite{luo2022demonstration} \\
\addlinespace
Architecture (\S\ref{sec:architecture}) &
geometry: decouples absorption path from collection path &
flexibility; large area; low cost &
surface-to-volume ratio multiplies traps; peak performance sacrificed &
core--shell arrays~\cite{lin2016core, ghamgosar2019core};
woven fibres~\cite{deng2023woven} \\
\bottomrule
\end{tabular}
\end{table}

Reading the matrix column by column: \emph{responsivity} is raised most reliably by junction
engineering and polarization coupling, amplified by architecture; \emph{speed}
favours thin-depletion junctions and Schottky contacts but is degraded by the deep
traps defect engineering introduces, so doping is double-edged; \emph{spectral
selectivity} is the domain of band-edge doping and wide-gap PEC devices;
\emph{stability} is the strength of all-oxide junctions and the weak point of
polarization (a transient, poling-dependent response) and PEC (a liquid electrolyte
to contain); and \emph{flexibility} and
\emph{low cost} are delivered by architecture and solution-processed routes, at
some expense of peak performance. These anticorrelations are not accidental:
responsivity and speed pull apart because the trap states that lengthen the
response also raise the low-frequency signal; peak performance and low cost pull
apart because the cleanest, highest-$V_\mathrm{bi}$ junctions demand vapour growth
while the cheap solution and PEC routes accept defective interfaces; and stability
sits opposite the two strongest boosters, a transient polarization response and a liquid
electrolyte. The matrix is thus less a scorecard than a map
of one tension, between the field strength one can build, the defect density
tolerated to build it cheaply, and the durability of whatever supplies the field.

Six device families show that the strongest results come from \emph{combining}
strategies rather than perfecting one. \emph{(i) Junction $+$ interface:} a
Ti$_3$C$_2$T$_x$ MXene/GaN van der Waals Schottky junction whose depletion depth
exceeds the UV absorption depth collects carriers with an external quantum
efficiency above 99\% at zero bias and gives stable output~\cite{song2021mxene};
interface quality decides how much of $V_\mathrm{bi}$ survives as photocurrent.
\emph{(ii) Polarization $+$ architecture:} patterning 2D ferroelectric
perovskite microbelts into ordered arrays on a flexible substrate improves
reproducibility over randomly placed single crystals, and tensile strain then adds
a piezo-phototronic gain to the ferro-pyro-phototronic response~\cite{guo2022patterned}.
\emph{(iii) Interface $+$ architecture:} a Ga$_2$O$_3$ nanorod-array PEC cell
pairs the near-zero-cost solid/liquid junction with a high-surface-area
morphology, giving low-cost solar-blind detection with no solid p--n
junction~\cite{chen2019photoelectrochemical}. \emph{(iv) Doping $+$ junction:}
Ga--In co-doped ZnO with an asymmetric Al--Au contact merges carrier-density
tuning with a work-function field, which alone makes the device
self-powered~\cite{tsay2022improved}. \emph{(v) Junction geometry, for speed:} a
2D--3D In$_2$O$_3$ homojunction speeds interfacial charge transfer in a PEC
photoanode~\cite{zhang2025accelerating}, while in a p--i--n stack the i-layer
thickness trades photogenerated carrier number against field strength, the best
zero-bias response appearing near 30~nm~\cite{chen2023deep}.
\emph{(vi) Doping $+$ interface:} Mg-doped $\alpha$-Ga$_2$O$_3$ nanorods
in a PEC cell form V$_\mathrm{O}$--Mg$_\mathrm{Ga}$ complexes that raise interfacial
catalytic activity and carrier transport, enough for a 5$\times$5 solar-blind
array used in underwater imaging~\cite{zhou2025doped}. Across
all six, no single lever is responsible; the skill lies in choosing which two or
three to combine for the target at hand.

A fully quantitative responsivity-versus-speed compilation would complement
Fig.~\ref{fig:matrix}, but the metrics that would populate it are defined
inconsistently across the literature (Section~\ref{sec:metrics}): responsivities
at unstated irradiance, shot-noise-only detectivities, ambiguous rise-time and
active-area definitions. Assembling it rigorously is a data-curation task in its own
right, so we leave it as a follow-up to be reported with explicit provenance
rather than from headline numbers here.

\begin{figure}[htbp]
\centering
\includegraphics[width=\linewidth]{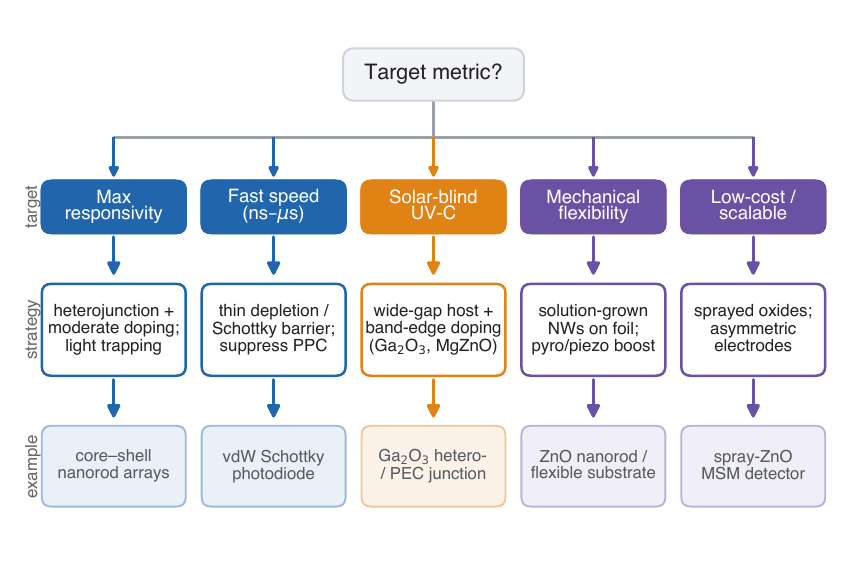}
\caption{Design-rule decision tree: each target metric maps to a recommended
strategy combination and a representative growth route and example system,
colour-coded by the dominant strategy family of Fig.~\ref{fig:umbrella}.}
\label{fig:tree}
\end{figure}

\section{Design rules and outlook}\label{sec:designrules}

\subsection{Design rules}

The matrix of Fig.~\ref{fig:matrix} can be read forwards, from a target metric to
a recommended strategy and growth route (Fig.~\ref{fig:tree}). For \emph{maximum
responsivity}, use a well-aligned heterojunction with moderate (not maximal)
doping and a light-trapping architecture; the doping is kept moderate because the
zero-bias optimum sits below the conductivity optimum (Section~\ref{sec:paradox}). For
\emph{fast response}, favour a thin depletion region or Schottky barrier and
actively suppress deep traps and persistent photoconductivity. \emph{Solar-blind
UV-C} points to a wide-gap host (Ga$_2$O$_3$, MgZnO) with band-edge doping, as a
heterojunction or a low-cost PEC cell. \emph{Mechanical flexibility} calls for
solution-grown nanostructures on a compliant substrate, where a polarization boost
can compensate a weaker junction. \emph{Low cost and scalability} favour
solution processing, such as spray-pyrolysed or hydrothermally grown oxides with
asymmetric electrodes. These are
heuristics, not guarantees: most real targets are multi-objective, and the value
of the tree is in making the dominant trade-off explicit before fabrication
begins.

The tree is a sequence of commitments. First the host and gap: solar-blind UV-C
forces a wider-gap oxide or an alloyed edge, fixing the achievable
$V_\mathrm{bi}$ and growth route. Second the field source: a solid junction for
stability, a polarization layer for pulsed targets, or a PEC cell where cost and
solar-blind selectivity outweigh liquid packaging. Only then does doping enter, as
a paradox-constrained \emph{tuning} step: set the carrier density to the
intermediate value balancing field against depletion width, confine heavy doping
to a contact layer, and alloy the band edge only within the range that keeps the host
phase~\cite{zhou2025mgzno}. Spray
pyrolysis and hydrothermal synthesis suit the doping, PEC, and architecture
families; epitaxy is the price of the cleanest junctions.

\subsection{Open problems}

Several bottlenecks recur across the families and define the near-term agenda.
First, \emph{persistent photoconductivity} and slow trap-mediated tails remain the
chief obstacle to fast zero-bias response, and they are governed by exactly the
native defects that doping perturbs, a defect-engineering problem, not a
new-material one. Second, \emph{p-type doping} of wide-gap oxides is still neither
reproducible nor stable, which is why self-powered oxide homojunctions remain rare
(Section~\ref{sec:doping}); a robust p-type ZnO or Ga$_2$O$_3$ would reshape the
field. Third, \emph{reproducibility and batch variation} in defect-engineered
films (sensitive to atmosphere, temperature, and precursor chemistry, and
especially severe in solution-based chemical growth) hinder translation of a
laboratory optimum into a manufacturable process, and are under-reported. Fourth, \emph{reporting consistency}: until detectivity, response
time, and active area are stated on a common
basis~\cite{wang2023characterize, dearquer2017solution}, cross-study comparison
and any data-driven optimization rest on unstable ground. This exposes a
characterization gap: deep-level transient spectroscopy and positron
annihilation remain rare, so the deep levels that govern zero-bias speed and
persistent photoconductivity are seldom probed with the techniques best able to
resolve them.

Fifth, \emph{operational stability is largely unmeasured}, and this is the gap
most likely to separate the published literature from a deployable device. Each
family carries its own degradation risk, from the ambient instability of halide
perovskites~\cite{veeramalai2021lead} to humidity, which raises leakage and dark
current in metal-oxide detectors~\cite{jain2024heterostructured}, yet the standard evidence
offered is a few minutes of on/off switching; the exceptions that report elevated-temperature operation or stable output over
repeated switching~\cite{sheoran2022zero, song2021mxene}
remain the minority. The asymmetry is stark: the physical
processes act over months, the measurements over minutes, and the conclusions are
drawn about years. Reporting cycle count, elapsed time, atmosphere, and
encapsulation as part of the performance claim (Section~\ref{sec:metrics}) costs
nothing beyond instrument time and would make the stability column of
Table~\ref{tab:families} an empirical rather than an inferred one.

Sixth, \emph{area scaling and uniformity} are almost never addressed. The
low-cost, large-area promise that motivates solution-processed self-powered
detectors~\cite{dearquer2017solution, jiang2024solution} is tested on
millimetre-scale devices, whereas the defect landscape that
governs zero-bias operation is exactly what varies across a sprayed or
hydrothermally grown film~\cite{nurfani2019substrate, nurfani2023molarity}. Because responsivity depends on where the field sits
relative to the absorption profile, spatial variation in thickness, grain size, or
vacancy density translates directly into pixel-to-pixel variation in an array, and
this cannot be calibrated away when the variation is also drift-prone. Reporting
device-to-device statistics across a substrate, rather than a champion device,
would do more to establish manufacturability than a further increment in peak
responsivity.

Two are concrete material-physics targets. Persistent photoconductivity is
attributed to metastable oxygen-vacancy configurations and slow re-adsorption of
surface oxygen; suppressing it means controlling a defect's charge-state
thermodynamics, and because it is measurable through the decay transient it should
be reported alongside the rise time. The p-type problem stems from deep acceptor
levels, self-compensation by native donors, and a low-lying valence band, so the
credible routes are valence-band engineering, defect-equilibrium control, and
p-type partner oxides as a stopgap~\cite{chen2025spiro, xu2024spraying}, not
brute-force doping. Both reduce to which point defect sits where in the gap, which
is why the near-absence of deep-level-resolving characterization lets them
persist.

Looking ahead, two directions stand out. Machine-learning-guided screening of
dopant--defect combinations suits the high-dimensional, trade-off-laden design
space mapped here, but (not yet applied systematically to self-powered oxide detectors) its value hinges on harmonized rather than headline training data:
the ceiling on statistical screening is label consistency, not the
algorithm~\cite{wang2023characterize}. Second, the step from a ``zero-bias''
device to a genuinely \emph{autonomous} system (the detector co-designed with on-chip energy storage and signal conditioning) remains largely unexplored for
UV oxides, and is where the cost and stability advantages catalogued here will
ultimately be tested. That is where a self-powered \emph{component} becomes a
self-powered \emph{sensor}.

\section{Conclusions}\label{sec:conclusions}

Operating a ZnO UV photodetector without external bias is not a property of any
one contact or mechanism but a design problem: building, then optimizing, the
internal asymmetry that separates and collects carriers at zero bias. Organizing
the field around ZnO by engineering strategy rather than by material or effect
makes that problem tractable. Junction engineering remains the default for
steady-state responsivity and stability; doping and defect control is the
cheapest, most versatile lever but carries the zero-bias paradox, the
reproducibility burden, and the p-type bottleneck; polarization coupling buys
transient responsivity at the cost of reporting ambiguity; interface and
photoelectrochemical engineering offers low-cost solar-blind detection bounded by
chemical stability; and architecture delivers flexibility and low cost at some
expense of peak performance. The strongest devices combine two or three levers,
and the strategy--metric matrix and decision tree presented here are meant to make
those combinations deliberate. Two messages stand out. First, the zero-bias
optimum is genuinely different from the biased optimum (most visibly for doping, where more is not better), and conflating the two has
slowed progress.
Second, the remaining bottlenecks are problems of defect control and rigorous,
consistent reporting, not of discovering new absorbers. Built on earth-abundant
oxides and low-temperature solution routes, self-powered UV detection is also an
inherently sustainable technology; realizing it will come from engineering the
asymmetry and the defect landscape with precision, and from measuring the result
without overstating it.

\section*{CRediT authorship contribution statement}
\textbf{Eka Nurfani:} Conceptualization, Visualization, Writing -- original draft.
\textbf{Keysya C. A. Putri:} Writing -- review \& editing.
\textbf{Tarmizi Taher:} Writing -- review \& editing.
\textbf{Aditya Rianjanu:} Writing -- review \& editing.
\textbf{Muhamad F. Arif:} Writing -- review \& editing.

\section*{Declaration of competing interest}
The authors declare no competing financial interests or personal relationships
that could have appeared to influence the work reported in this paper.

\section*{Funding}
This research received no specific grant from any funding agency in the public,
commercial, or not-for-profit sectors.

\section*{Acknowledgements}
The authors thank Muhammad A.\ K.\ Purbayanto (Faculty of Mechatronics, Warsaw
University of Technology) for a careful critical reading of the manuscript and
constructive suggestions, and colleagues at Institut Teknologi Sumatera (ITERA)
for helpful discussions during the preparation of this review.

\section*{Data availability}
No new experimental data were generated or analysed in this review; all
performance values discussed are taken from the cited literature.

\section*{Declaration of generative AI and AI-assisted technologies in the
writing process}
During the preparation of this work the authors used Claude (Anthropic) in order
to support literature screening and reference curation, check cited statements
against the source texts, and edit the language. After using this tool, the
authors reviewed and edited the content as needed and take full responsibility
for the content of the published article.

\printbibliography

\end{document}